\documentclass[11pt,letterpaper,twocolumn]{article}

\usepackage{cvpr}
\usepackage{times}
\usepackage{epsfig}
\usepackage{graphicx}
\usepackage{amsmath}
\usepackage{amssymb}

\usepackage[pagebackref=true,breaklinks=true,letterpaper=true,colorlinks,bookmarks=false]{hyperref}

\cvprfinalcopy 

\ifcvprfinal\fi
\begin{document}

\title{A Quantitative Analysis of Localized Robustness of MYCN in Neuroblastoma}

\author{Romeil Sandhu$^1$, Sarah Tannenbaum $^2$,  Daniel Diolaiti$^2$, Alberto Ambesi-Impiombato$^2$ \\ Andrew Kung$^2$, Allen Tannenbaum$^1$\\
{\small [1] Departments of Computer Science and Applied Mathematics, Stony Brook University}\\
{\small [2] Division of Pediatric Hematology/Oncology/Stem Cell Transplantation, Columbia University Medical Center}}

\maketitle
\thispagestyle{empty}

\begin{abstract}
The amplification of the gene MYCN (V-myc myelocytomatosis viral-valeted oncogene, neuroblastoma derived) has been a well-documented indicator for poor prognosis in neuroblastoma, a childhood cancer.  Unfortunately, there has been limited success in understanding MYCN functionality in the landscape of neuroblastoma and more importantly, given that MYCN has been deemed ``undruggable,'' the need to potentially illuminate key opportunities that indirectly target MYCN is of great interest.  To this end, this work employs an emerging quantitative technique from network science, namely network curvature, to quantify the biological robustness of MYCN and its surrounding neighborhood.   In particular, when amplified in Stage IV cancer, MYCN exhibits higher curvature (more robust) than those samples with under expressed MYCN levels.  When examining the surrounding neighborhood, the above argument still holds for network curvature, but is lost when only analyzing differential expression - a common technique amongst oncologists and computational/molecular biologists.  This finding points to the problem (and possible solution) of drug targeting in the context of complexity and indirect cell signaling affects that have often been obfuscated through traditional techniques.
\end{abstract}

\section{Introduction}
Neuroblastoma, the most common extra-cranial tumor in childhood, is an embryonic tumor that is derived from the neural crest \cite{Leiqi}.  It can occur anywhere along the sympathetic nervous system, most commonly the adrenal glands in addition to the neck, chest, abdomen, pelvis, and spine, and the clinical manifestations are varied depending on location and severity.  The prognosis for low and intermediate risk neuroblastoma is excellent, whereas that for high risk disease remains around 40\% despite intensive chemotherapy and myeloblative therapy with stem cell transplantation \cite{Matthay}.  Other novel approaches have emerged in the treatment of high risk disease; however drug resistance continues to be the main obstacle in finding a cure for this disease.  One of the main indicators for high risk neuroblastoma is MYCN amplification, found in approximately 25\% of cases and correlated with poor prognosis \cite{Huang}.  Although directly targeting MYCN is not currently feasible, current efforts are underway to inhibit targets directly or indirectly associated with the gene.  Motivated by such work, this note focuses to better understand MYCN in a quantitative manner via biological robustness.  In turn, targets of opportunity that indirectly counter-effect MYCN (e.g., specific pathways) may be uncovered and furthered in the clinical setting with respect to drug development.

To do so, we use the geometric network method proposed in our previous work \cite{our_nature,ewing}.  These papers are centered around several key concepts.  Firstly, cell signaling cascades within neuroblastoma can be viewed as a complex biological network (weighted graph) whereby the nodes in the network represent genes and the edges characterize the ``strength'' of interaction between such genes. Secondly, the underlying network may be alternatively viewed as a discrete statistical manifold for which those edge weights represent a one-step random walk.  In doing so, we are then able to show that Ricci curvature (from geometry) is positively correlated with a networks robustness \cite{our_nature}.  Biologically speaking, this has allowed us to differentiate cancer tissue networks from their normal counterparts on several tumor types, i.e., we were able to quantitatively show that cancer is more robust than normal tissue (given a precise definition of robustness as given in Section \ref{sec:curve_robust}).  Using the same methodology, we were further able to show that drug resistant samples in Ewing Sarcoma were more robust than untreated and drug sensitive samples and that drug sensitive samples exhibited fragility as compared to untreated samples \cite{ewing}.   Nevertheless, while in these previous works, we took a more global view in our analysis of the cancer network, in the present work we considered a more localized analysis centered around MYCN for neuroblastoma.

The remainder of the present note is outlined as follows:  In the next section, we revisit the concept of Ricci curvature in general and for the discrete case of graphs.  We further expound upon the flexibility of providing a local proxy for robustness by introducing discrete scalar curvature that is capable of accounting for inherent cell-signaling complexity.    Section~\ref{sec:Results} then presents localized curvature results to illuminate that MYCN amplification is robust and that it also promotes robustness in the neighboring signaling region.  This is particularly compelling as differential expression is unable to elucidate such representative information in this same neighborhood further stressing the need for network science principles when performing a quantitive analysis. We conclude this brief note with Section \ref{sec:FutureWork} discussing future work.

\section{Preliminaries}
In this section, we provide background on geometric concept of curvature.  We note that much of this background can be similarly found in previous works \cite{our_nature,ewing,our_science} and is simply provided here to make this note more self-contained.

\subsection{Introduction to Ricci Curvature}
We begin with introducing some background on Ricci curvature \cite{ollivier_intro}; we refer the reader to \cite{docarmo} for all the rigorous mathematical details. Accordingly, let $X$ be a Riemannian manifold (the generalization of a smooth surface valid in any dimension). One can naturally measure distances on $X$ and thus define the length of a given curve.  \emph{Geodesics} are curves that locally are the shortest distance between two given points on the manifold $X$, and are essential to defining curvature.

\begin{figure*}[!t]
\begin{center}
\includegraphics[height=5cm]{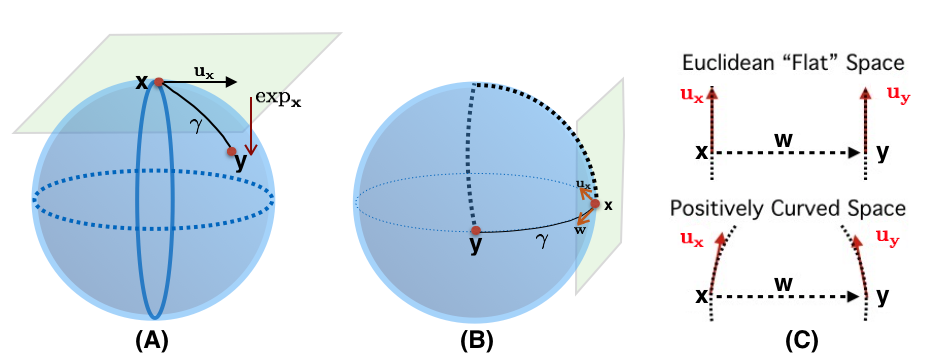}
\caption{An Intuitive Understanding of Curvature. (A) Given a point $x$ on a Riemannian manifold, one has a corresponding tangent plane for which a set tangent vectors lie upon (i.e., $u_x$).  Locally, there exists a geodesic curve $\gamma$ that for two close enough points, the distance traveled along this curve represents the distance between such points.  If one was to travel along the tangent vector, the endpoint (which lies on the tangent space) may be mapped back to the manifold - this mapping is known as the exponential map.  (B) Using this basic geometric understanding, one may define the notion of curvature by comparing two tangent vectors at two respective points connected by a geodesic in a given direction, which is shown to be $w$ in the above example.  (C) From this, curvature is the deviation of such tangent vectors from a Euclidean space.  In the simplest Euclidean space, such endpoints of those vectors form a rectangle.  In the case of the sphere, we have positive curvature whereby geodesics along $u_x$ and $u_y$  converge.  Similarly, in the case of negative curvature, they diverge.}
\label{fig:fig2}
\end{center}
\end{figure*}

More specifically, let $x\in X$,  let  $T_x$ denote the tangent space at $x$ with $u_x, w \in T_x$  denoting two orthogonal unit vectors. Then if we move along the geodesic curve $\gamma$ at $x$  in the direction of $w$ in an infinitesimal manner, we let $y\in X$ be the endpoint of the traversal. A pictorial representation of this is given in Figure \ref{fig:fig2}A-B. The transversal is carried out via parallel transport, that roughly allows us to connect the geometry of nearby points in a canonical manner \cite{docarmo}. We denote by $u_y \in T_y$, the result of the parallel transporting of $u_x$ to the point $y$. On the plane, this would exactly correspond to moving a given vector in a parallel manner along a straight line (the geodesics of Euclidean space). We call $u_y$  the \emph{parallel transport} of $u_x$ .

On a curved space, geodesics defined along $u_x$ and $u_y$ (denoted by $\exp_{x} tu_x$ and $\exp_{y}tu_y$,  respectively) may converge towards one another (positive curvature) or diverge from one another (negative curvature); see Figure \ref{fig:fig2}C. The is called {\em geodesic deviation}.  With this in mind, we are able to define Ricci curvature via sectional curvature \cite{ollivier_intro,docarmo}. Again for $u_y$ the parallel transport of $u:=u_x$ from point $x$  to $y$  in the direction $w$, we have for sufficiently small $\epsilon, \delta > 0$,
\begin{equation}
d(\exp_{x} \epsilon u_x,\!\exp_{y} \epsilon u_y)\!\!=\!\delta(1-\frac{\epsilon^{2}}{2} K(u,w) + O(\epsilon^3 + \epsilon^2\delta)).
\end{equation}
The term $K(u,w)$ denotes the \emph{sectional curvature} at $x$ in the tangent plane $(u,w)$.

Classical Ricci curvature is defined by averaging $K(u,w)$ over all directions in the $(u,w)$ tangent plane. Of course, the above construction is defined in the continuous setting to provide an intuitive insight and we will now turn our attention to the discrete setting that will be needed for graphs and how we can measure the fragility/complexity of MYCN within neuroblastoma.

\subsection{Graph Ricci Curvature}

Given we are working on a discrete graph/network where ordinary notions of smoothness are not applicable, we need to extend notions of Ricci curvature to such a setting. We should note that there are number of possibilities \cite{LV,Mass,Zhou,Romania,McC97,tetali,Sturm,Gromov} that are geared towards extending the notion of Ricci curvature to more general metric spaces. Here, we employ a clever notion of a Ricci curvature, due to Ollivier \cite{Ollivier,Oll_markov,ollivier_intro}, based on a synthetic coarse geometric approach.  We will call this notion {\em Ollivier-Ricci curvature} and Ricci curvature will always be taken to be in this sense.

In order to define the Ollivier-Ricci curvature, we will need to define the \emph{Wasserstein 1-distance} (also called the \emph{Earth Mover's Distance}) \cite{Villani2,Villani3} on a discrete metric measure space  $X=\{x_1,...,x_n\}$. Let $\mu_1$  and $\mu_2$ denote two distributions having the same total mass, and denote by $d(x,y)$ the distance between $x,y\in X$  (for graphs, taken as the hop metric).  Then $W_1(\mu_1,\mu_2)$ is be defined as follows \cite{Rubner}:
\begin{equation}
W_1(\mu_1,\mu_2) = \min_{\mu} \sum_{i,j=1}^{n} d(x_i,x_j) \mu(x_i,x_j)
\end{equation}
where $\mu(x,y)$ is a \emph{coupling} (i.e., distribution on $X\times X$) subject to the following constraints:
\begin{align}
\label{constraints}
\mu(x,y) &\geq 0 \hspace{9mm} \forall x,y \in X \\
\sum_{i=1}^{n} \mu(x, x_i) &= \mu_1(x) \hspace{3mm} \forall x\in X \\
\sum_{i=1}^{n} \mu(x_i,y) &= \mu_2(y) \hspace{3mm} \forall y \in X.
\end{align}
The cost above finds the optimal coupling of moving mass defined by distribution $\mu_1$ to $\mu_2$ with minimal ``work.''  Very importantly, the Wasserstein 1-distance may be computed as a linear program allowing for an efficient, highly parallelizable algorithm.

We can now define the\emph{ Ollivier-Ricci curvature}. The intuition underpinning the approach is motivated from the observation (in the classical continuous case) that the distance between two small (geodesic) balls is less than the distance between their centers on a positively curved space (and greater than the distance between the centers on a negatively curved one).  Accordingly, if we let $(X,d)$  be metric space equipped with a family of probability measures $\{\mu_x : x \in X\}$, we define the \emph{Ollivier-Ricci curvature} $\kappa (x,y)$ along the geodesic connecting nodes $x$ and $y$ via
\begin{equation} \label{OR}
W_1(\mu_x,\mu_y) = (1-\kappa(x,y))d(x,y),
\end{equation}
where $W_1$ is the Wasserstein distance defined above \cite{Villani2,Villani3,Evans1989,Tannenbaum2010signals,Givens} and $d$ is the distance on $X$. For the case of weighted graphs, we set
\begin{eqnarray} d_x &=& \sum_y w_{xy}\\
\mu_x(y)&:=& \frac{w_{xy}}{d_x} ,\end{eqnarray}
where $d_x$ is the sum taken over all neighbors of node $x$ and where $w_{xy}$  denotes the weight of an edge connecting node $x$ and node $y$  ($w_{xy}=0$ if $d(x,y)\geq 2$). The measure  $\mu_x$ may be regarded as the distribution of a one-step random walk starting from $x$, with the weight $w_{xy}$ quantifying the strength of interaction between nodal components or the diffusivity across the corresponding link (edge).

\begin{figure}[!t]
\begin{center}
\includegraphics[height=8cm]{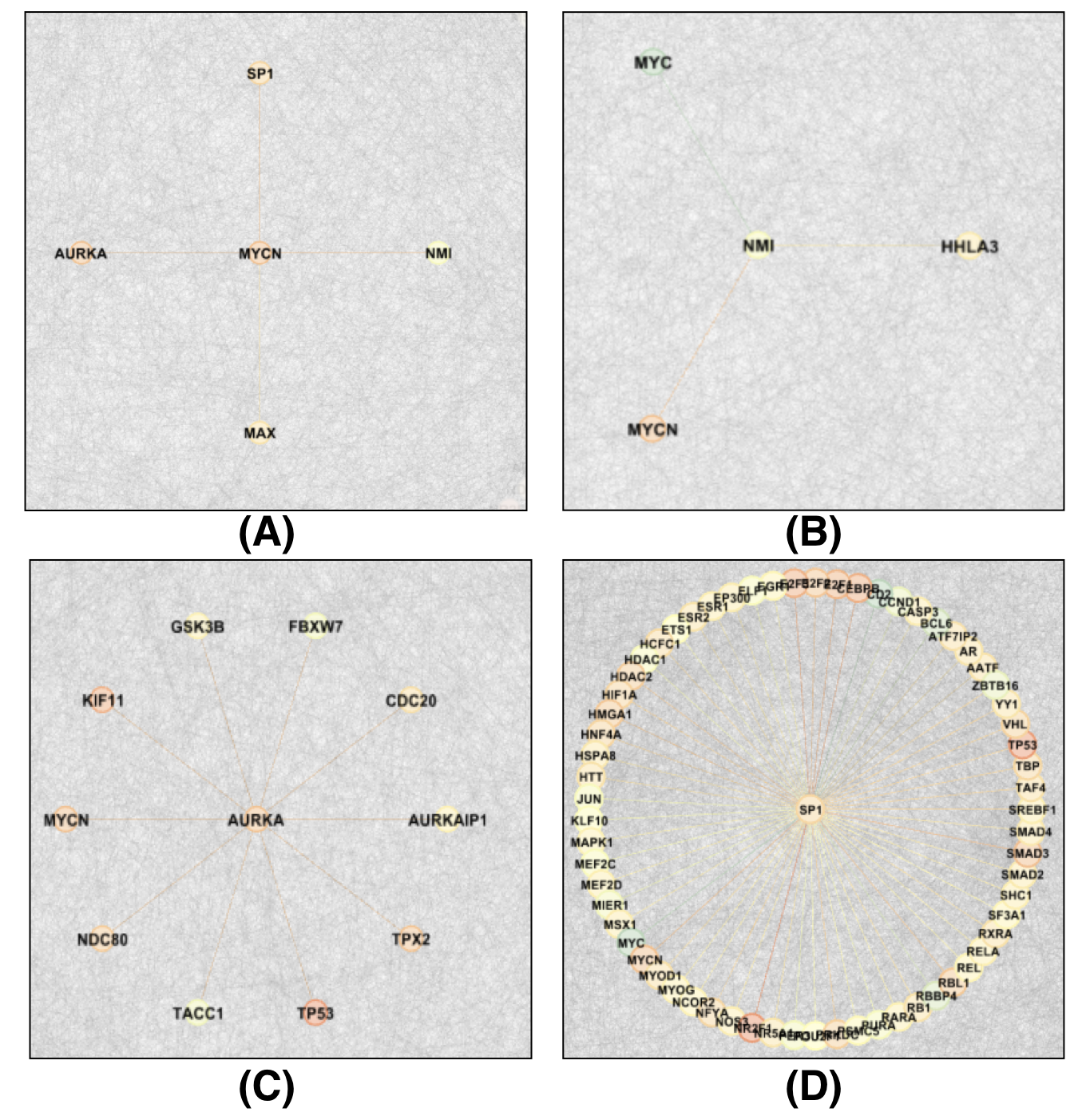}
\caption{The constructed biological network of MYCN neighborhood in neurbolastoma.  (A) Direct interactions of MYCN: NMI, MAX, AURKA, SP1 (B) Direct interactions of NMI.  (C) Direct interactions of AURKA.  (D)  Direct interactions of SP1.}
\label{fig:fig3}
\end{center}
\end{figure}

\begin{table*}[!t]
\begin{center}
\includegraphics[height=2.5cm]{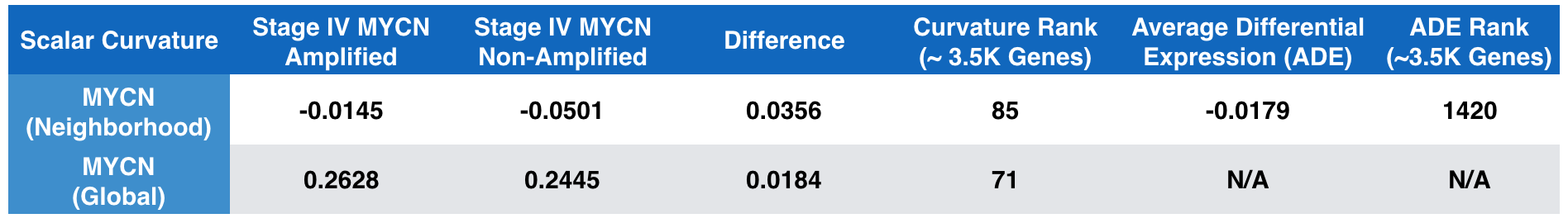}
\caption{This table presents scalar curvature results for Stage IV amplified and non-amplified MYCN cases.  One can see that curvature illustrates MYCN robustness.  Note I:  The average differential expression of MYCN in Amplified vs Non Amplified: 0.7070 (Rank 261 out of $\approx$ 3.1 K genes).  Note II: ``Neighborhood'' is defined as any gene that can be reached to MYCN in two ``hops.''}
\label{fig:fig4}
\end{center}
\end{table*}

\subsection{Curvature and robustness} \label{sec:curve_robust}

We summarize here the central relationship of curvature and robustness, namely their positive correlation.
We express this as $$\Delta R \times \Delta Ric \ge 0,$$  where $\Delta R$ denotes change in robustness and $\Delta Ric$ change in curvature. This is strongly related to the \emph{Fluctuation Theorem}, which shows that changes in entropy is positively correlated to changes robustness \cite{dem}, and this relationship has been exploited in \cite{West} for cancer as well. The relation between entropy and curvature is described in \cite{LV} using deep results in the theory of optimal transport.

We just note here that {\em robustness} is characterized by the ability of a system to functionally adapt to changes in the environment. The formal definition is based
on the theory of large deviations
\cite{varadhan}. One considers random fluctuations of a given network that result in perturbations of some observable. We let $p_\epsilon (t)$ denote the
probability that the mean deviates by more than $\epsilon$ from the original (unperturbed) value at time $t$. Since $p_\epsilon(t) \rightarrow 0$ as $t \rightarrow \infty,$ we want
to measure its relative rate, that is, we set
$$R := \lim_{t \rightarrow \infty} (-\frac{1}{t} \log p_\epsilon(t)).$$ Therefore, large $R$ means not much deviation and small $R$ large deviations.

Because of the positive correlation of curvature and robustness, one can use curvature as a proxy for robustness. This gives a major advantage since curvature may be computed via a linear program while the computation for robustness may be quite challenging.

\subsection{Weighted Graph Scalar Curvature}

Based on our preceding discussion, the above measure of Ollivier-Ricci curvature provides a proxy for edge (gene-to-gene interaction) robustness.  This is a local property capable of analyzing a pathway that connects \emph{any} two given genes or proteins in a biological network.  While this allows us to examine interactions involving MYCN, we also seek to understand MYCN as a stand alone gene, in the context of robustness, compared to other genes involved in our network.  The important caveat here is a single gene (MYCN), when measured in a given neighborhood (or globally), must account for \emph{all direct and indirect interactions}.  This is a problem of complexity--how can we quantitatively account for such enumerated pathways.

We approach the above issue via the related notion of \emph{scalar curvature}, which for intuitive purposes, can be considered as the average of aforementioned Ricci curvature.  Specifically, scalar curvature at a given node $x$ can be defined on a discrete graph as a weighted contraction with respect to the probability distribution $\mu_x(y)$ as $S(x):=\sum_y \mu_x(y)\kappa(x,y)$.  However, as mentioned, we would like to provide a metric that is capable of measuring network robustness in a localized region with respect to a particular gene and in turn, must measure all pathways that interact with that gene.  As follows, we alternatively define local-to-global ``scalar'' curvature $S_{lg}(x)$ as follows:
\begin{equation}
\label{eq:SC}
S_{lg}(x):=\frac{1}{N_{\hat{y}}}\sum_{\hat{y}} \kappa(x,y)
\end{equation}
where $\hat{y}$ are those nodes that fall within a particular $\xi$-geodesic neighborhood (i.e., $\hat{y} = \{y\in X : d(x,y)\leq \xi\}$) and $N_y$ are the number nodes $\hat{y}$.  One could also weight the above contraction with respect to the measure.

\section{Results}
\label{sec:Results}
Prior work \cite{our_nature} has shown that cancer networks exhibits a higher degree of robustness as compared to normal tissue networks. While we highlighted the importance of understanding biological pathway fragility, much of the analysis was conducted at a macroscopic level over several cancer types.  Here, we provide a localized analysis of a single cancer study (neuroblastoma) with an even more specific focus on a known poor prognosis indictor (amplification of MYCN).  To the best of our knowledge, this is the first result to show that MYCN exhibits a higher degree of biological robustness in Stage IV samples when amplified as opposed to non-amplified Stage IV samples.   However, before doing so, we first present details on the obtained data and corresponding network construction.

\subsection{Data and Network Construction}
In this study, we obtained raw gene expression data from ArrayExpress (url: http://www.ebi. ac.uk/arrayexpress, study: E-MTAB-1781).  In particular, 709 neuroblastoma samples were available including Stage I, Stage II, Stage IV tumor grades.  There was a total of $161$ Stage IV samples for which MYCN was considered ``non-amplified'' and $95$ samples that were considered ``amplified.'' In addition to this, $355$ Stage I/II samples were available and used.

With regards to constructing the underlying interactome (topology of the graph/network), we utilized three separate databases:  Human Protein Reference Database (HPRD), Human Interactome Project (LIT) and String v10.  In particular, to ensure a highly accurate neighborhood construction of the MYCN neighborhood (e.g., those direct and primary indirect interactions), we first sought out all possible interactions that were listed in the HPRD and LIT database.  Then, such interactions were re-examined and reconstructed by \emph{String v10} / Gene-Cards, taking only those interactions with a high confidence ($>.7$) to ``clean'' the neighborhood.  The final result was further verified by our collarborators in the Division of Pediatric Hematology/Oncology/Stem Cell Transplantation at Columbia University, and the result of this neighborhood can be seen in Figure~\ref{fig:fig3}.  For the remaining interactions, we  used HPRD as our primary source for the underlying network.  Using the above raw expression data, we were able to compute correlation between any two given genes and these values served as the weights in our corresponding network.  Note:  Given that correlation ranges from [-1,1], we utilized an affine transformation to ensure positive weights (i.e., $\hat{c}_{xy}= \frac{1}{2}(c_{xy}+1)$).  Accordingly, we built four separate biological networks composed of $\approx$ 3.5K genes:  MYCN Amplified Stage IV, MYCN Non-Amplified Stage IV, Stage IV (irrespective of MYCN amplification), and Stage I/II (irrespective of MYCN amplification).

\subsection{Robustness of MYCN}
We begin by computing Ricci curvature $\kappa(x,y)$ on Stage IV samples of both amplified and non-amplified MYCN cases.  In particular, while it is known that MYCN amplification in ``high risk'' (Stage IV) patients signify a poor outcome in terms of patient survivability, there has been limited success in being able to properly target MYCN and to further expound upon the functionality of MYCN with respect to neuroblastoma.  This said, using equation (\ref{eq:SC}), we are  able to compute the changes in curvature (signifying changes in robustness) on such samples.  These results are seen in Table \ref{fig:fig4}.  Interestingly, the neighborhood of MYCN, defined to be only those interactions that are direct and primary indirect interactions, exhibits higher scalar curvature (more robust) than the non-amplified samples.  In contrast, if one were to compute the the average differential expression of this neighborhood, information regarding MYCN significance is seemingly ``lost.'' The significance of this result can be seen in terms of gene ranking with respect to scalar curvature and differential expression.  Examining MYCN more globally, i.e., taking \emph{all} interactions involving MYCN, we see that curvature is able to aptly characterize its significance (via rank) as compared to simply differential expression.  This quantification is consistent with previous thinking and current on-going clinical work.

In addition, we also examined sample that comprise Stage IV and Stage I/II data, regardless of MYCN amplification.  These results can be seen in Table \ref{fig:fig5}.  While MYCN is still considered to be ``robust'' in Stage IV as compared to Stage I/II, the results are much less pronounced.  Biologically speaking, this would make sense as cure rates for Stage I/II are considerably high and do not need generally involve targeting MYCN, i.e., MYCN is not considered to be a key factor contributing to the robustness of neuroblastoma.   Even further and in Stage IV, MYCN has only been the focus in those samples for which it has been amplified.  Indeed, the considerable risk that has been noted in clinical outcomes primarily relies only on MYCN amplification in Stage IV cancer.  Together, these two, albeit short studies, provide an interesting path forward to understand how MYCN functions in the context of drug resistance.

\section{Future Work}
\label{sec:FutureWork}
To the best of our knowledge, this work provides the first quantitive localized analysis with respect to quantitive (biological) robustness of MYCN in neuroblastoma. In particular, through a network science approach, we were able to show that those samples in Stage IV neuroblastoma for which MYCN is amplified, also exhibits higher network curvature.  Through existing  work \cite{our_nature}, which suggests that increases in network curvature is positively correlated to increases in network robustness, we were able to deduce that such findings point to the robustness of MYCN for ``high risk'' patients.  This is particularly compelling in that, while differential expression is unable to elucidate the importance of these indirect cell signaling affects, our network curvature-based approach is able to do so.  From this, the next step, which will be a subject of future research, is to quantify which feedback pathways contribute to MYCN robustness and more importantly, which drug targets may be considered (e.g., AURKA) that will decrease such robustness.
\begin{table*}[!t]
\begin{center}
\includegraphics[height=2.5cm]{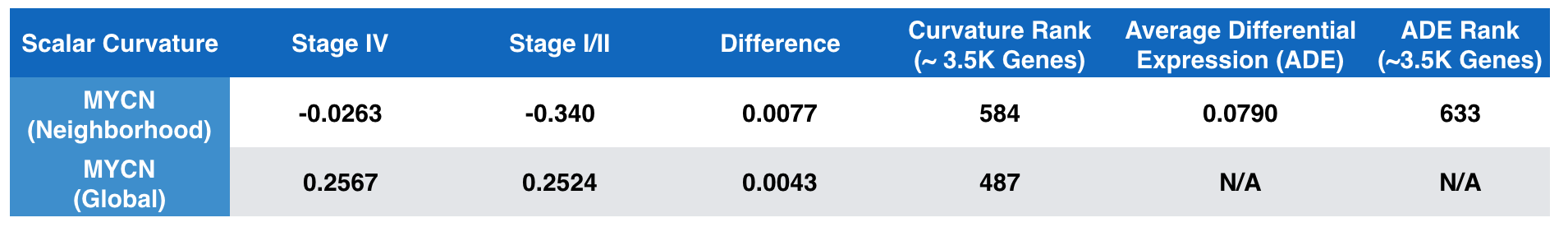}
\caption{This table presents MYCN scalar curvature results for Stage IV samples compared to that of Stage I/II, irregardless of MYCN amplification.  One can see that curvature illustrates MYCN robustness; however, the results are less pronounced compared to Stage IV study.  Note I:  The average differential expression of MYCN in Amplified vs Non Amplified: 0.6557 (Rank 229 out of $\approx$ 3.5 K genes).  Note II: ``Neighborhood'' is defined as any gene that can be reached to MYCN in two ``hops.''}
\label{fig:fig5}
\end{center}
\end{table*}

\section*{Acknowledgements}
This project was supported by in part by grants from the National Center for Research Resources (P41-
RR-013218) and the National Institute of Biomedical Imaging and Bioengineering (P41-EB-015902)
of the National Institutes of Health. This work was also supported by NIH grant
1U24CA18092401A1 as well as AFOSR grants FA9550-12-1-0319 and FA9550-15-1-0045.

\end{document}